\RequirePackage{fix-cm}
\documentclass{svjour3}       
\smartqed  \usepackage{graphicx}
\usepackage[utf8]{inputenc}
\usepackage{lineno,hyperref,stfloats}
\usepackage{amsmath,amssymb,bm}
\usepackage{graphicx}
\usepackage{makecell}
\usepackage{natbib}
\usepackage[dvipsnames]{xcolor}

\providecommand{\R}{\mathbb{R}}

\providecommand{\mb}[1]{\mathbf{#1}}
\usepackage[normalem]{ulem}
\usepackage[colorinlistoftodos,backgroundcolor=white,linecolor=Gray]{todonotes}

\allowdisplaybreaks

\begin{document}

\title{Structural identifiability analysis of PDEs: \\A case study in continuous age-structured epidemic models}

\titlerunning{Structural identifiability analysis of PDEs}  

\author{Marissa Renardy         \and
        Denise Kirschner \and Marisa Eisenberg 
}

\institute{M. Renardy \at
              University of Michigan Medical School, Department of Microbiology and Immunology \\
              \email{renardy@umich.edu}           
           \and
           D. Kirschner \at
              University of Michigan Medical School, Department of Microbiology and Immunology \\
          \and
          M. Eisenberg \at
          University of Michigan, Department of Epidemiology \\
          University of Michigan, Department of Mathematics \\
          \email{marisae@umich.edu}
}

\date{Received: date / Accepted: date}

\maketitle

\begin{abstract}
Computational and mathematical models rely heavily on estimated parameter values for model development. Identifiability analysis determines how well the parameters of a model can be estimated from experimental data. Identifiability analysis is crucial for interpreting and determining confidence in model parameter values and to provide biologically relevant predictions. Structural identifiability analysis, in which one assumes data to be noiseless and arbitrarily fine-grained, has been extensively studied in the context of ordinary differential equation (ODE)  models, but has not yet been widely explored for age-structured partial differential equation (PDE) models. These models present additional difficulties due to increased number of variables and partial derivatives as well as the presence of boundary conditions. In this work, we establish a pipeline for structural identifiability analysis of age-structured PDE models using a differential algebra framework and derive identifiability results for specific age-structured models. We use epidemic models to demonstrate this framework because of their wide-spread use in many different diseases and for the corresponding parallel work previously done for ODEs. In our application of the identifiability analysis pipeline, we focus on a Susceptible-Exposed-Infected model for which we compare identifiability results for a PDE and corresponding ODE system and explore effects of age-dependent parameters on identifiability. We also show how practical identifiability analysis can be applied in this example.
\end{abstract}

\section{Introduction}
Identifiability analysis addresses the question of whether the parameters of a mathematical model can be uniquely identified, given observed data. There are two general types of identifiability analysis: structural (assuming data to be noiseless and arbitrarily fine-grained) and practical (using real data). These analyses are crucial to interpreting biologically relevant predictions from computational and mathematical models that rely on parameter values that are estimated from experimental datasets. Previous studies have shown that unidentifiable models can lead to vastly different predictions for different parameter values that are able to produce the same observed data \citep{Kao18}. In the context of biological models, parameter identifiability gives increased confidence in model predictions that are based on estimated parameters. This is particularly relevant for predictions applied to topics such as diseases, treatment outcomes, vaccination efficacy, and future infection dynamics. Specifically, the use of epidemiological models has become increasingly necessary and useful to predicting disease outcomes, as we have seen in the case of COVID-19, and these efforts should be as accurate and confident as possible to produce the most useful predictions. Thus, we focus our work in the context of epidemic models due to their use in many diseases and to compare with corresponding parallel work previously done for ODEs regarding structural identifiability.

Many theoretical results and algorithms for structural identifiability of linear and non-linear ordinary differential equations (ODEs) models have been derived \citep{miao2011,Audoly2001,Ljung1994,Pohjanpalo1978,Chappell1990,Cobelli1980,chis2011structural} and have been applied to study an array of diseases. There are models for the spread of cholera \citep{Eisenberg13}, vector-borne diseases \citep{Kao18,Zhu18,Tuncer16}, and general infectious diseases \citep{Tuncer18,Evans05}. Methods for structural identifiability are often analytical in nature (e.g. the commonly used differential algebra approach \citep{Audoly2001,Ljung1994}), but may also be numerical, or some blend of the two \citep{Jacquez1985,Jacquez1990,Raue2009}. 

ODE models are not always sufficient for representing disease dynamics when, for example, age is an important factor. In many diseases, different age groups may respond differently to diseases or disease interventions based on differences in susceptibility \citep{Gardner80}, mortality \citep{Chavez91}, and contact rates \citep{Mossong08,Glasser12}. Age preferences have been shown to have a significant effect on mixing patterns that drive disease spread \citep{DelValle07}. This type of information can be incorporated into a mathematical model for epidemics, for example, by explicitly representing age-structured distributions of the susceptible and infected subpopulations. Age-structured models have been used to evaluate and optimize age-targeted vaccination strategies for several diseases \citep{Harris19,Keeling11,Hao19,Chavez98,Shim06}, and it has been shown that population heterogeneity and non-random mixing can have a substantial effect on the prediction of disease outbreaks and herd immunity \citep{Glasser16,Feng15}. 

In a continuous setting, age structure is introduced by creating state variables that are functions of both age and time, and also by introducing transport terms to represent constant aging of a population. These model formulations result in PDEs, rather than ODEs, making them more biologically accurate; however, such models also typically more complex to analyze. There are other PDE formulations of epidemic models such as age of infection models \citep{Thieme93,Inaba04} or spatially explicit models \citep{Bertuzzo09,Wu2008}. The focus of this paper will be specifically on age-structured models. While there are alternative approaches to incorporating age structure, such as to use compartmentalization to stratify populations \citep{Chow11,Ferguson96} or to use a stochastic discrete representation such as an agent-based model \citep{Ajelli10,Bian04}, we focus here on the PDE formulation. Importantly, the issues of both structural and practical parameter identifiability for these types of models still needs to be elucidated.

While structural identifiability analysis has been extensively applied in the context of ODE epidemic models, it has not been widely explored for age-structured PDE models. PDE models present additional difficulties due to increased number of variables and derivatives as well as the presence of boundary conditions. Identifiability of age-structured models was first considered by \citet{Perasso11} for a simple SI-type model and was again considered by \citet{Perasso16} for a population dynamics model including only birth and death. 
These analytical results have not yet been extended to more comprehensive epidemic models, though some identifiability results have been shown numerically \citep{Tuncer16}. For these types of problems, modelers have relied on practical identifiability analyses which are dependent on either available data or data that has been synthesized for the purpose of analysis.

In this paper, we derive analytical identifiability results for more complex age-structured epidemic models using a differential algebra framework that is simpler to apply than the previous (aforementioned) previous approaches. We apply this methodology to a compartmental epidemic Susceptible-Exposed-Infected (SEI) model both with and without age structure explicitly modeled, and we compare the identifiability results under different conditions such as the presence of immigration in the system or for age-dependent model parameters. To corroborate our structural identifiability results, we also perform a practical identifiability analysis for one version of the model. While we focus on age-structured models, the methods we present herein can be applied to other PDE formulations, with certain assumptions about smoothness and uniqueness of solutions.

\section{Methods}
\subsection{Identifiability framework}
To consider parameter identifiability of a model, we must first define the model structure. From an identifiability or parameter estimation perspective, a model is composed of two pieces: (1) a state model or set of state equations, which describe the underlying system of interest, and (2) a measurement model or set of \emph{output equations}, which describe the observation or measurement of interest (i.e., the quantities of interest for which one either has or plans to collect data). In this study, we assume the following general first-order partial differential equation (PDE) model structure, which is commonly used in age-structured PDE models in biology and health:
\begin{equation}\label{eq:model}
\begin{aligned}
\frac{\partial \mathbf{x}}{\partial t} + \frac{\partial \mathbf{x}}{\partial a} &= f(\mathbf{x}, t,a,\mathbf{p}, \mathbf{u})\\
\mathbf{y} &= g(\mathbf{x}, t,a,\mathbf{p})
\end{aligned}
\end{equation}
where $\mathbf{x}$ is a vector of state variables describing the system of interest, $\mathbf{p}$ is the vector of parameters, $t$ and $a$ are two independent variables. Here we consider time and age as the independent variables, although the approach given here will also work for more general variables, as well as for additional derivatives of further independent variables on the left hand side. Additionally, $\mathbf{y}$ represents the vector of $n_y$ model outputs (measured variables) and $\mathbf{u}$ represents any known inputs (e.g. forcing functions) to the system, if they exist. Finally, $f$ and $g$ are rational functions of the model variables and parameters such that the model can be written equivalently as differential polynomials by clearing denominators. 

In structural identifiability analysis, it is assumed that the inputs and outputs are known (observed) perfectly -- in this case, we assume that $u$, and $y$ are known for all times and all ages and that there is no noise in the data. Similarly, the independent variables $t$ and $a$ are assumed to be known. Thus, the resulting identifiability results can be interpreted as the upper limit of what can possibly be identified with perfect data collection methods. However, structural identifiability is a necessary condition for identifiability in the realistic data case, making it an important first step in the parameter estimation process. 

A model such as~\eqref{eq:model} can be thought of as a map from parameter space (e.g. for $\mathbf{p} \in \R^{n_p}$) to output space (the space of trajectories for $\mathbf{y}$), often termed the model map $\Phi: \mathbf{p} \mapsto \mathbf{y}$. With this framing, the structural identifiability question becomes that of whether or not the model map $\Phi$ is injective. In particular, we define structural identifiability as follows:

\begin{definition} For a model of the form~\eqref{eq:model}, an individual parameter $p$ in $\mb{p}$ is \emph{uniquely (or globally) structurally identifiable} if for almost every value $\mb{p}^*$ and almost all initial conditions, the equation $\mb{y}(\mb{x},t,a,\mb{p}^*) = \mb{y}(\mb{x},t,a,\mb{p})$ implies $p = p^*$.  A parameter $p$ is said to be \emph{non-uniquely (or locally) structurally identifiable} if for almost any $\mb{p}^*$ and almost all initial conditions, the equation $\mb{y}(\mb{x},t,a,\mb{p}^*) = \mb{y}(\mb{x},t,a,\mb{p})$ implies that $p$ has a finite number of solutions.  \end{definition}

\begin{definition}  Similarly, a model of the form~\eqref{eq:model} is said to be \emph{uniquely} (respectively, \emph{non-uniquely}) \emph{structurally identifiable}  for a given choice of output $\mb{y}$ if every parameter is uniquely (respectively, non-uniquely) structurally identifiable, i.e. for almost every value $\mb{p}^*$ and almost all initial conditions, the equation $\mb{y}(\mb{x},t,a,\mb{p}^*) = \mb{y}(\mb{x},t,a,\mb{p})$ has only one solution, $\bf p = p^*$ (respectively, finitely many solutions).  Equivalently, a model is uniquely structurally identifiable for a given output if and only if the map $\Phi$ is injective almost everywhere, i.e. if there exists a unique set of parameter values $\mb{p}^*$ which yields a given trajectory $\mb{y}(\mb{x},t,a,\mb{p}^*)$ almost everywhere. \end{definition}

We note that because some specific parameter values or initial conditions may render an otherwise identifiable model unidentifiable (e.g. if all parameters or initial conditions are zero), we define structural identifiability generically, i.e., for almost all parameter values and initial conditions \citep{Audoly2001,Saccomani2003}.

\subsection{Overview of the differential algebra approach}
To determine structural identifiability of a model with defined outputs, we extend the differential algebra approach \citep{Ljung1994,Audoly2001} to the age-structured PDE case. This approach relies on converting the model system into a system of input-output equations, that is, a set of $n_y$ monic polynomial equations in terms of the known variables $\mathbf{y}$, $\mathbf{u}$, and their derivatives, with rational coefficients in the parameters $\mathbf{p}$. The input-output equations represent an implicit form of the model map, and as a set of differential equations, they are input-output equivalent to the original system. This means that for the same inputs and parameter values, the equations generate the same output trajectories \citep{Eisenberg19arxiv}. We then consider the map from the input parameter space to the coefficients of the resulting monic polynomial. 

A key insight of the differential algebra approach in the rational ODE case is that injectivity of the model map can be evaluated by considering the the coefficient map, i.e. the map from parameters to coefficients of the input-output equations \citep{Audoly2001,Ljung1994}. In the rational ODE case, the coefficient map typically provides complete identifiability information for the model. If this map is injective, then the model is structurally identifiable. If the model is unidentifiable, this map determines the identifiable combinations of parameters. For example, while some model parameters may not be independently identifiable, the product or sum of two or more model parameters may be identifiable. This may enable the model to be reparameterized in a way that allows the model to become structurally identifiable \citep{Chappell98,Cole10}. It should be noted that a set of identifiable combinations is not necessarily unique -- there may be many sets of parameter combinations that are equivalent. For example, the sets of identifiable combinations $\{ab, bc\}$ and $\{ab,a/c\}$ are equivalent since $(ab)/(bc)=a/c$.

The differential algebra approach for ODE models typically uses characteristic sets to calculate the input-output equations \citep{Audoly2001,Ljung1994}. Characteristic sets are a differential analog of the Gr\"{o}bner basis \citep{ritt1950differential} based on differential polynomial psuedodivision. For more details on the characteristic set approach, the reader is referred to \citep{Audoly2001,Ljung1994,Saccomani2003,hong2020global,Eisenberg19arxiv}. 

This methodology has previously been applied to ODE systems, where derivatives are taken with respect to only one variable. In this work, we generalize this methodology to PDE systems of the form in~\eqref{eq:model}; in particular, we consider age-structured systems where derivatives are taken with respect to both time and age. We show that the same framework can be applied to these systems, with the additional consideration of boundary conditions. In age-structured systems, a boundary condition must be specified at $a=0$, usually in the form of a Dirichlet boundary condition. Thus, in addition to performing identifiability analysis on the model equations, one must also consider the equations evaluated at $a=0$ to explicitly include information at the boundary. This is analogous to methods for determining identifiability of initial conditions in ODE systems \citep{Saccomani2003}. 

\subsection{Extending the differential algebra approach to age structured PDE models}
In this section, we discuss our generalization of the differential algebra approach to the PDE case. For PDE models, the primacy of the differential ideal generated by the model and measurement equation differential polynomials is not necessarily guaranteed (at least, to our knowledge it has not been shown to be the case), meaning that the characteristic set may not be unique. Thus, instead of a characteristic set based approach, we will use a simpler, substitution-based approach based on previous work by \citet{Eisenberg19arxiv} (similar to that used in \citet{Eisenberg13}). In this approach, we use basic substitution (which, as long as one is careful not to divide by quantities that could be zero, etc., does not change the underlying solutions to the system) to eliminate the unobserved variables, resulting in a monic, reduced system of differential equations which is input-output equivalent to the original system \citep{Eisenberg19arxiv}. This input-output equivalent system has equivalent identifiability properties to the original, allowing us to use it to evaluate identifiability of PDE systems. 

As is typical in ODE systems, once we have a set of input-output equations (i.e. monic differential polynomials only in terms of the known variables $\bf y, u$ and their derivatives with respect to either independent variable and the parameters $\bf p$), we assume that our observed trajectory contains sufficiently many algebraically independent points that we can successfully identify the coefficients of the input-output equations. In other words, from a given observed solution of $\bf y$ over age and time, we assume the observed solution contains an arbitrary number of distinct points at which the input-output equations can be evaluated to solve for the input-output coefficients (in particular, more distinct points than the number of coefficients). In practice most trajectories meet this criterion, although some care must be taken to avoid systems or measure zero sets of initial conditions for which the solutions are at steady state, or have key parameters equal to zero, etc. \citep{hong2020global}. With the coefficient-identifiability assumption \citep{Eisenberg19arxiv}, the rest of the differential algebra approach follows as with the ODE case, and is described below. 

Procedurally, the process of performing identifiability analysis for PDEs is almost identical to the ODE case, though substitutions may be made more complicated by the presence of an additional derivative. To obtain input-output equations from model equations, a typical first step is to solve the output equations for state variables and, using substitution, eliminate these state variables and their derivatives from the model equations. For simplicity, all nonzero terms in the model equations are moved to the right-hand side. Remaining state variables can then be eliminated by solving one or more of the model equations. The particular substitutions required will of course vary depending on the model and output equations. The goal is to obtain a system of polynomial equations in terms of the outputs and their derivatives, with all state variables eliminated, whose coefficients are in terms of the model parameters. In the case that square roots arise, for example, they must be eliminated by multiplying by the conjugate. 

Once input-output polynomial equations are obtained, we divide each equation by the coefficient of its leading term to obtain a monic polynomial (see below for details about the ranking of terms).  From these monic polynomials, we obtain the set of polynomial coefficients that defines the coefficient map. This map, as in the ODE case, can be used to determine complete identifiability information for the model equations.

Note that thus far, we have not considered any information about the boundary conditions. Analysis of the boundary conditions is a key component of identifiability analysis for PDEs that does not occur in the ODE case. In the case of age-structured systems, there is a single boundary condition at $a=0$. Thus, we treat the boundary condition as similar to an initial condition for ODEs \citep{Saccomani2003}. If information is known about the boundary conditions, such as values or functional forms, this information can be substituted into the model equations evaluated at $a=0$ to obtain \emph{boundary condition equations}. Since the values of model outputs are known at $a=0$, the same identifiability analysis as above can be applied to boundary condition equations. This may reveal additional identifiability information about the model or any parameters that are present in the boundary conditions.

We note that, as in the initial condition case, if identifiability is evaluated without considering boundary conditions, this implicitly assumes that the boundary conditions are generic and unknown. However, there are some aspects of the boundary conditions which are more subtle than the ODE case of initial conditions, and may warrant further exploration, such as that boundary conditions still may vary over time. Exploring how different boundary condition types may affect identifiability would be an important direction for future work.

\begin{figure}[htbp!]
    \centering
    \includegraphics[width=85mm]{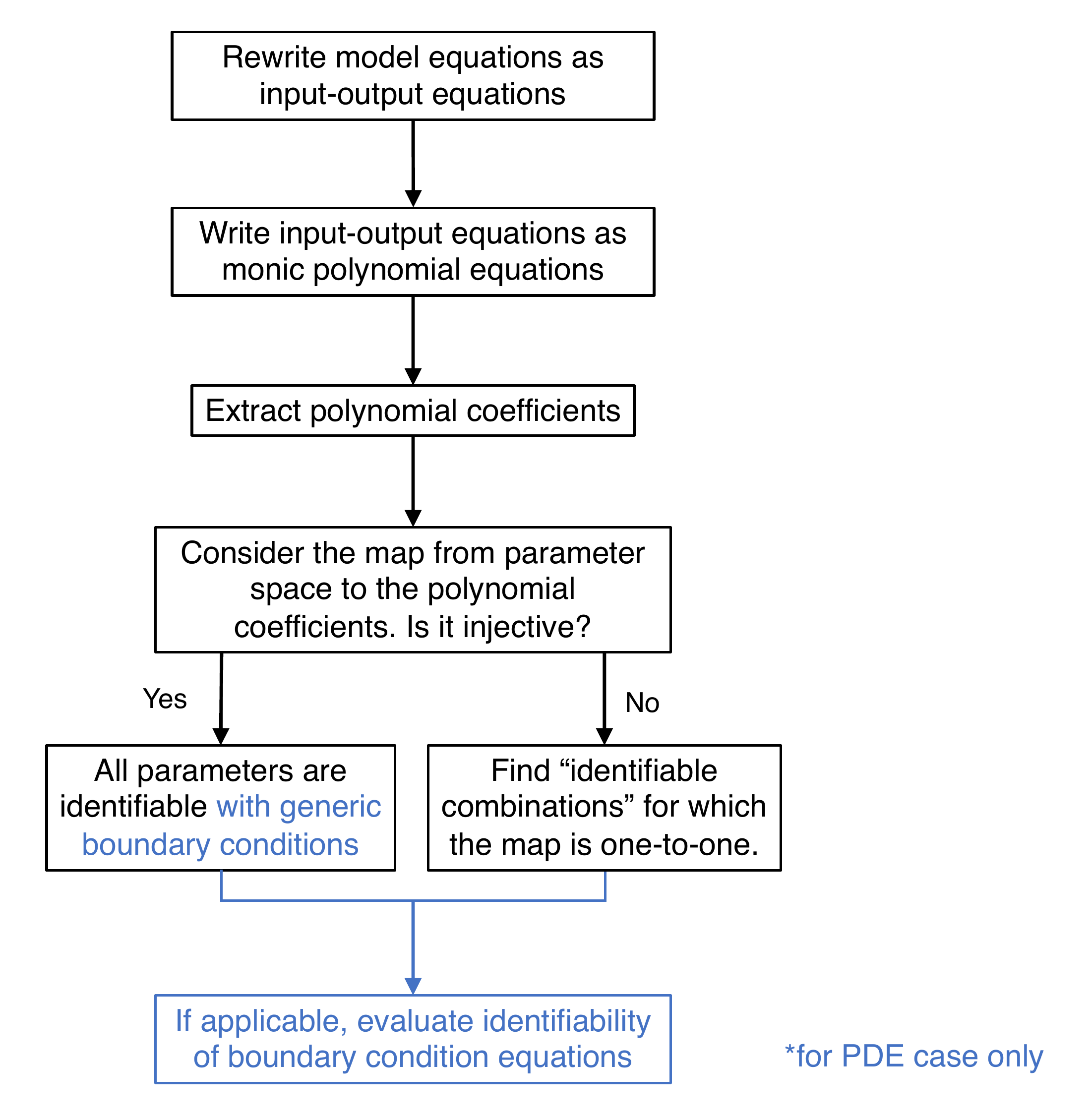}
    \caption{Flowchart describing the methodology for identifiability analysis of ODE and age-structured PDE models based on using a differential algebra approach.}
    \label{fig:flowchart}
\end{figure}

Our procedure for performing structural identifiability analysis, which can be applied to both ODE and age-structured PDE systems, is outlined in Figure \ref{fig:flowchart}.

\subsection{Practical identifiability analysis}\label{sec:practical}
Practical identifiability was evaluated using the Fisher Information Matrix (FIM) and profile likelihoods \citep{Jacquez1990,Raue2009,Venzon88,eisenberg2014determining}. Here, we follow the methodology laid out in \citep{eisenberg2014determining}.

For ODE models, the FIM is computed by $F=X^T X$, where $X$ is the sensitivity matrix defined by $X_{(i,j)}=\frac{\partial y}{\partial p_i}(t_j)$ for parameters $p_1, ..., p_n$ and time points $t_1, ..., t_m$. For an age structured system, we define $F = X^T X$ where the sensitivity matrix is defined by $X_{(i,j)} = \frac{\partial y}{\partial p_i}((a,t)_j)$ where $(a,t)_j$ for $j=1,...,m$ are pairs of age and time points that form a grid. Here, $y$ denotes the observable model output. For multiple model outputs (e.g., $y_1$ and $y_2$), the matrix $X$ can be defined by concatenation (i.e., $X = [X_1; X_2]$). The rank of $F$ indicates the number of identifiable parameters and parameter combinations. If $F$ is singular, the model is structurally unidentifiable; if $F$ is close to singular, the model may be practically unidentifiable. The FIM can also be used to determine which parameter subsets are identifiable or unidentifiable \citep{eisenberg2014determining}.

We compute profile likelihoods by fixing a particular parameter $p_i$ and fitting all remaining parameters $p_{j \neq i}$ to maximize a likelihood function. We define the likelihood by assuming constant, normally distributed measurement error with standard deviation equal to 10\% of the mean of the data, and mean equal to the model trajectory. Maximizing the likelihood function is equivalent to minimizing a cost function based on the negative log likelihood (in this case equivalent to least squares). This is performed at many values across the range of values for $p_i$. The profile likelihood for $p_i$ is given by the maximum values of the likelihood function across the range of values for $p_i$. If the profile likelihood is flat, $p_i$ is structurally unidentifiable; if it is sufficiently shallow, $p_i$ is practically unidentifiable. For more details on profile likelihood approaches to identifiability, the reader is referred to \citep{Raue2009,Venzon88}.

In our analyses herein, we solved the age-structured PDEs numerically using Method of Lines, using a first order finite difference approximation of the spatial derivative, and using Matlab's ODE solver {\it ode45}.

\section{SEI model for TB epidemiology}

In several case studies below, we consider a simple mathematical model of the epidemiology of tuberculosis (TB). TB is transmitted via the aerosolized bacterium {\it Mycobacterium tuberculosis} from person to person via lung inhalation. TB is unique from many infectious diseases in that most people develop a {\it latent} infection, meaning that they present no symptoms and cannot transmit the disease, and the infection can remain latent for an extended period of time. Thus, our model consists of three mutually exclusive compartments: Susceptible (S), Exposed/latent (E), and Infectious (I). Through contact with infectious individuals, susceptible individuals (called susceptibles) may progress to the exposed class (i.e. latent infection) or the infectious class (i.e. primary infection). This is slightly different from the standard SEI model, in that some individuals bypass the exposed class and immediately become infectious, but is similar to what has been used to model tuberculosis epidemics previously \citep{Ozcaglar12,Guzzetta11,Renardy19,Vynnycky97}. Individuals who are exposed/latent may progress to the infectious class where they can transmit bacteria. We assume all individuals are born into the susceptible class, and each compartment has a compartment-specific death rate. 

\begin{figure}[htbp!]
    \centering
    \includegraphics[width=65mm]{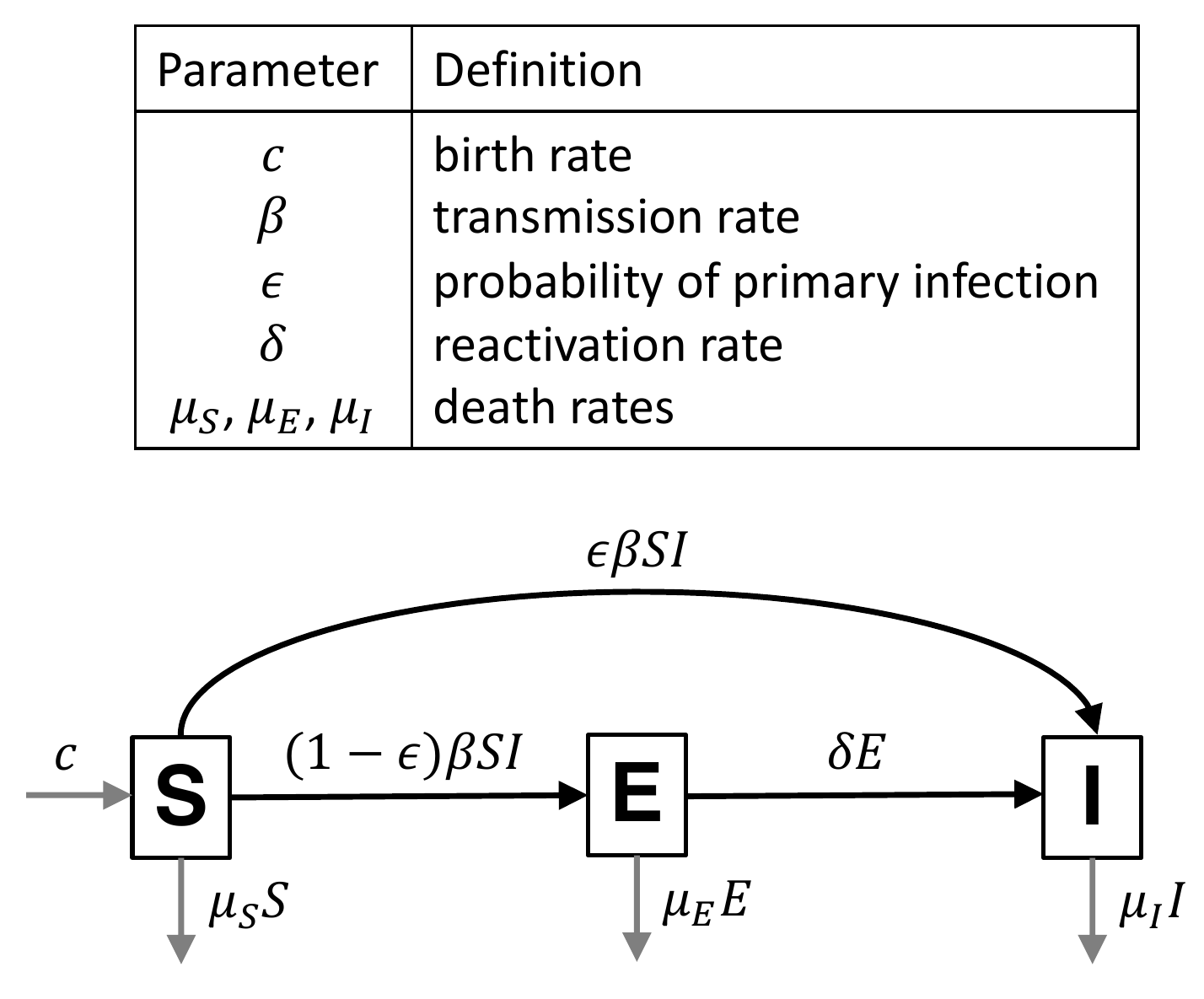}
    \caption{SEI diagram and parameter definitions for a mathematical model of TB epidemiology. {\it S} represents the susceptible population, {\it E} represents the exposed/latent population, and {\it I} represents the actively infectious population. Arrows describe flow between mutually exclusive compartments.}
    \label{fig:SEI_diagram}
\end{figure}

A schematic for this model is shown in Figure \ref{fig:SEI_diagram}. This schematic can be translated into a mathematical or computational model by representing it as a system of ordinary or partial differential equations (as has been done by, e.g., \citet{Guzzetta11,Renardy19,Vynnycky97}), a stochastic model, or an individual-based model (as has been done by, e.g., \citet{Guzzetta11,Renardy20}). Here, we will first consider the model as a system of ODEs and then extend the results to age-structured PDEs.

\section{Case study 1: Density-dependent transmission}
For simplicity, we first demonstrate the identifiability analysis framework on a model with density-dependent transmission, i.e. transmission is proportional to $SI$ rather than $S\frac{I}{N}$. In this case, the per-individual rate of contact increases as the population density increases. This is not necessarily a realistic model for TB, since TB is often spread only among close contacts, but simplifies the differential algebra for the purposes of demonstration.

In order to perform parameter identifiability analysis, we must first define what the outputs (i.e., the quantities for which we could obtain reliable, biologically relevant time series data) of the specific model are. For all of the following examples, we will assume that we can observe a fraction of the exposed individuals, i.e. $y_E = k_E E$ where $k_E$ is the reporting/detection rate. This type of data could be obtained by, for example, contact tracing to identify individuals who have been exposed to TB disease, or testing a proportion of the population for exposure. This is a useful practice even in areas with a low TB burden such as the US, since detection and treatment of latent infection is crucial for TB elimination \citep{LoBue17}. Since latent disease is asymptomatic, it is highly unlikely that all exposed/latent individuals will be identified, and thus we assume instead that only a fraction $k_E$ of the exposed individuals are identified. 

In addition, we may be able to observe all or some of either all active infections (prevalence) or new active infections (incidence). The relevant output function depends on the reporting and surveillance data for the region of interest. For example, if all infectious cases are monitored and reported then these data would represent total prevalence. If not all cases are detected due to misdiagnosis, for example, then the data would represent only a fraction of the total prevalence. If new infectious cases are reported upon diagnosis but then lost to follow-up, then incidence is an observable output while prevalence is not. Again, since it is possible that not all infected individuals will be diagnosed, we may only be able to observe a fraction of incidence. We present here results for observing prevalence over time, but the identifiability analysis of model parameters for incidence would follow similarly.

We assume that $y_I = k_I I$, i.e. that we are able to obtain data for a fraction of the total number of infections at any given time. We assume no other information about the system parameters or initial conditions. 

\subsection{ODE SEI model with density-dependent transmission} \label{sec:ODE-dens}
Though identifiability analysis for ODE systems is well established, we include the technical details here for completeness and for comparison when we move to the PDE system. The model can be written as a system of ODEs:
\begin{equation}
\label{eq:ODE_SEI}
\begin{aligned}
\frac{dS}{dt} &= c-\beta SI - \mu_S S \\
\frac{dE}{dt} &= (1-\epsilon)\beta SI - \delta E - \mu_E E \\
\frac{dI}{dt} &= \epsilon \beta SI + \delta E - \mu_I I
\end{aligned}
\end{equation}

We begin by rewriting the system of equations \eqref{eq:ODE_SEI} in terms of the outputs $y_E$ and $y_I$. For ease of notation, we adopt the notation $y'=\frac{dy}{dt}$. Using the substitutions $E=\frac{y_E}{k_E}$ and $I=\frac{y_I}{k_I}$, we obtain the following equations:
\begin{equation*}
\begin{aligned}
S' &= c-\beta S \frac{y_I}{k_I} - \mu_S S \\
\frac{y_E'}{k_E} &= (1-\epsilon)\beta S \frac{y_I}{k_I} - \delta \frac{y_E}{k_E} - \mu_E \frac{y_E}{k_E} \\
\frac{y_I'}{k_I} &= \epsilon \beta S \frac{y_I}{k_I} + \delta \frac{y_E}{k_E} - \mu_I y_I.
\end{aligned}
\end{equation*}
It now remains to eliminate $S$ from these equations so that we obtain equations in terms of solely the model parameters and outputs. This can be done by solving for $S$ in the equation for $\frac{dy_I}{dt}$:
\[ S = \frac{k_E (y_I'+\mu_I y_I)-\delta  k_I y_E}{\beta  k_E \epsilon  y_I}. \]

This expression can then be substituted into the remaining two equations to obtain two input-output equations that are rational equations in terms of the outputs and their derivatives. By multiplying by a common denominator and moving terms to one side, these input-output equations can be rewritten as the following polynomial equations.
\begin{equation*}
\begin{aligned}
0 &= k_I^2 \delta \mu_S y_E y_I + (c k_E k_I \beta \epsilon - k_E k_I \mu_I \mu_S) y_I^2 + k_I \beta \delta y_E y_I^2  \\
& \quad\quad - k_E \beta \mu_I y_I^3 + k_I^2 \delta y_I y_E' - k_I^2 \delta y_E y_I' - k_I k_E \mu_S y_I y_I' \\
&\quad\quad - k_E \beta y_I^2 y_I' + k_E k_I y_I'^2 - k_E k_I y_I y_I'' \\
0 &= (-k_I \delta - k_I \epsilon \mu_E) y_E + (k_E \mu_I - k_E \epsilon \mu_I) y_I - k_I \epsilon y_E' \\
&\quad\quad + (k_E - k_E \epsilon) y_I'
\end{aligned}
\end{equation*}
Dividing the first equation by $k_E k_I$ and the second equation by $k_E-k_E\epsilon$ to form monic polynomials, we get the following set of polynomial coefficients.
\begin{equation*}
\begin{aligned}
   \left\{-1,1,\frac{\beta }{k_I},-\frac{\delta  k_I}{k_E},\frac{\delta  k_I}{k_E},-\frac{\beta  \delta }{k_E},\frac{k_I \epsilon }{k_E (\epsilon -1)},\frac{k_I(\delta +\mu_E \epsilon )}{k_E (\epsilon -1)}, \right. \\
   \left. \frac{\beta  \mu_I}{k_I},\mu_I,\mu_S,-\frac{\delta  k_I \mu_S}{k_E},\mu_I \mu_S-\beta  c \epsilon \right\}
\end{aligned}
\end{equation*}

We consider this set of coefficients as a function of the model parameters, and we seek to determine whether this map is injective. If it is not, then we consider whether it is injective as a function of some parameter combinations. We find that $\mu_S$ and $\mu_I$ are structurally identifiable. Through simplification and substitution, we find that other identifiable parameter combinations are
\[ \left\{\frac{\beta}{k_I}, \beta c \epsilon, \frac{\beta\delta}{k_E},  \frac{\delta(-1+\epsilon)}{\epsilon}, \mu_E+\delta \right\}. \]
If we further assume that $\mu_S=\mu_E$, then $\delta$ and $\epsilon$ also become identifiable while $\beta$, $c$, $k_E$, and $k_I$ remain unidentifiable. However, the following combinations are identifiable: $\{ \frac{\beta}{k_E}, \frac{\beta}{k_I}, \beta c\}$. Consequently, if any one of $\beta$, $c$, $k_E$, or $k_I$ is known, then the rest become identifiable.

\subsection{Age-structured PDE model with density-dependent transmission and constant parameters}
\label{sec:PDE_constant-dens}

Now that identifiability results are established for the ODE representation of the SEI model \ref{eq:ODE_SEI}, we extend these results to a more complex PDE model that captures the same epidemiological setting (SEI), but includes age structure. We incorporate age structure into the model by adding a transport term to capture population aging. To simplify the model, we assume that transmission occurs between individuals of the same age, so that the transmission term is proportional to $S(t,a)I(t,a)$ for any given age $a$ and time $t$. While this is a significant simplification, data has shown that people tend to interact primarily with others in their age group \citep{Mossong08}. The model thus becomes
\begin{equation}\label{eq:PDE_model-dens}
\begin{aligned}
\frac{\partial S}{\partial t}+\frac{\partial S}{\partial a} &= -\beta SI - \mu_S S \\
\frac{\partial E}{\partial t}+\frac{\partial E}{\partial a} &= (1-\epsilon)\beta SI - \delta E - \mu_E E \\
\frac{\partial I}{\partial t}+\frac{\partial I}{\partial a} &= \epsilon \beta SI + \delta E - \mu_I I
\end{aligned}
\end{equation}
for $t>0$ and $a \in [0,\infty)$. Note that all state variables are now functions of both time and age. The boundary conditions are $S(t,0)=c$, $E(t,0)=I(t,0)=0$. Note that the birth rate is now incorporated into the boundary condition rather than into the differential equation for $S$. We begin by assuming that model parameters are constants, independent of age. 

\subsubsection{Identifiability of model equations with generic boundary conditions}

We first consider identifiability of the model parameters without explicitly accounting for the presence of boundary conditions, meaning that we assume there is no {\it a priori} knowledge about the boundary conditions and that they cannot be observed. We call this case {\it generic} boundary conditions. To perform the identifiability analysis, we utilize the same framework as we did for the above ODE system analysis. That is, we rewrite the model system as a set of polynomial equations in terms of the outputs and their partial derivatives, and then consider the map from parameter space to the polynomial coefficients. 

We adopt the notation $y^{(i,j)}=\frac{\partial^{i+j} y}{\partial t^i \partial a^j}$. Using the substitutions $E=\frac{y_E}{k_E}$ and $I=\frac{y_I}{k_I}$, as in the ODE case, we obtain the following equations:
\begin{equation*}
\begin{aligned}
S^{(1,0)}+S^{(0,1)} &= -\beta S \frac{y_I}{k_I} - \mu_S S \\
\frac{y_E^{(0,1)}}{k_E}+\frac{y_E^{(1,0)}}{k_E} &= (1-\epsilon)\beta S \frac{y_I}{k_I} - \delta \frac{y_E}{k_E} - \mu_E \frac{y_E}{k_E} \\
\frac{y_I^{(0,1)}}{k_I}+\frac{y_I^{(1,0)}}{k_I} &= \epsilon \beta S \frac{y_I}{k_I} + \delta \frac{y_E}{k_E} - \mu_I y_I.
\end{aligned}
\end{equation*}
We now eliminate $S$ to obtain equations in terms of only the model parameters and outputs. This can be done by solving for $S$ in the third equation:
\[ S = \frac{k_E \left(y_I^{(0,1)}+y_I^{(1,0)}+\mu_I y_I\right)-\delta  k_I y_E}{\beta  k_E \epsilon  y_I}. \]
As before, this expression can then be substituted into the remaining two equations to obtain two input-output equations that are rational equations of the outputs and their derivatives. By multiplying by a common denominator and moving terms to one side, these input-output equations can be rewritten as the following polynomial equations.
\begin{equation*}
\begin{aligned}
0 &= -k_E k_I \mu_S y_I^{(0,1)} y_I-k_E k_I \mu_S y_I^{(1,0)} y_I-k_E k_I y_I^{(0,2)} y_I \\
&\quad\quad -2 k_E k_I y_I^{(1,1)} y_I-k_E k_I y_I^{(2,0)} y_I+k_E k_I (y_I^{(0,1)})^2 \\
&\quad\quad +k_E k_I (y_I^{(1,0)})^2+2 k_E k_I y_I^{(0,1)} y_I^{(1,0)}-k_E k_I \mu_I \mu_S y_I^2 \\
&\quad\quad -\beta  k_E y_I^{(0,1)} y_I^2-\beta  k_E y_I^{(1,0)}y_I^2-\beta k_E \mu_I y_I^3 \\
&\quad\quad +\delta  k_I^2 y_E^{(0,1)} y_I+\delta  k_I^2 y_E^{(1,0)} y_I-\delta  k_I^2 y_E y_I^{(0,1)} \\
&\quad\quad -\delta  k_I^2 y_E y_I^{(1,0)}+\delta  k_I^2 \mu_S y_E y_I+\beta  \delta k_I y_E y_I^2 \\
0 &= -k_E \epsilon  y_I^{(0,1)}-k_E \epsilon  y_I^{(1,0)}+k_E y_I^{(0,1)}+k_E y_I^{(1,0)} \\
&\quad\quad +k_E \mu_I y_I-k_E \mu_I \epsilon  y_I-k_I \epsilon y_E^{(0,1)}-k_I \epsilon  y_E^{(1,0)}\\
&\quad\quad -\delta k_I y_E -k_I \mu_E \epsilon y_E
\end{aligned}
\end{equation*}
We choose derivatives with respect to time to be higher ranked than those with respect to age for our ranking of terms within the polynomial, and let $y_I$ be higher ranked than $y_E$, yielding the ranking $y_E < y_I< y_E^{(0,1)} < y_I^{(0,1)} < y_E^{(1,0)} < y_I^{(1,0)} < y_E^{(0,2)} < y_I^{(0,2)} < y_E^{(1,1)} < y_I^{(1,1)} < ... $. Dividing the first equation by $k_E k_I$ and the second equation by $k_E-k_E\epsilon$ to form monic polynomials, we get the following set of polynomial coefficients.
\begin{equation*}
\begin{aligned}
    \left\{ -2, -1, 1,2,\frac{\beta}{k_I}, -\frac{k_I\delta}{k_E}, \frac{k_I\delta}{k_E},\frac{k_I \epsilon }{k_E (\epsilon -1)},\frac{k_I (\delta +\mu_E \epsilon )}{k_E (\epsilon -1)}, \right. \\
    \left.\frac{\beta  \mu_I}{k_I},\mu_I,\mu_S,-\frac{\delta  k_I \mu_S}{k_E},\mu_I \mu_S-\frac{\beta  \delta }{k_E}\right\}
\end{aligned}
\end{equation*}

We find that, as was the case for the ODE model, $\mu_S$, and $\mu_I$ are structurally identifiable. Through simplification and substitution, we find that other identifiable parameter combinations are
\[ \left\{\frac{\beta}{k_I}, \frac{\delta k_I}{k_E}, \frac{\delta(\epsilon-1)}{\epsilon}, \mu_E+\delta \right\}. \]
If $\mu_S=\mu_E$ then $\delta$ and $\epsilon$ become identifiable as well as the combinations $\left\{ \frac{\beta}{k_E}, \frac{\beta}{k_I} \right\}$.

\subsubsection{Incorporating non-generic boundary conditions}
In our model, the boundary conditions are in fact not generic, but rather are known to be $S(t,0) = c, E(t,0)=I(t,0)=0$, where $c$ represents the birth rate. To determine the identifiability of the boundary condition parameter $c$, we perform the same analysis as above but with the model equations evaluated at $a=0$ (boundary condition equations). We let the boundary conditions for $E$ and $I$ be generic to avoid reducing equations to zero. Then, noting that $S(t,0)$ is constant and thus $\frac{\partial S}{\partial t}(t,0) = 0$, the relevant equations are
\begin{equation}\label{eq:PDE_boundary}
\small{
\begin{aligned}
\frac{\partial S}{\partial a}(t,0) &= -\beta cI(t,0) - \mu_S c \\
\frac{\partial E}{\partial t}(t,0)+\frac{\partial E}{\partial a}(t,0) &= (1-\epsilon)\beta cI(t,0) - \delta E(t,0)- \mu_E E(t,0) \\
\frac{\partial I}{\partial t}(t,0)+\frac{\partial I}{\partial a}(t,0) &= \epsilon \beta cI(t,0) + \delta E(t,0) - \mu_I I(t,0)
\end{aligned}
}
\end{equation}

By transforming the system \eqref{eq:PDE_boundary} to a set of polynomial equations we obtain the following polynomial coefficients:
\begin{equation*}
    \left\{ 1, -\frac{k_I \delta}{k_E}, -\frac{c k_E \beta - c k_E \beta \epsilon}{k_I}, \delta+\mu_E, \mu_I-c \beta \epsilon \right\}
\end{equation*}
Since $\mu_I$ is identifiable from the model equations, we thus find that $\beta c \epsilon$ is an identifiable combination, which is consistent with the ODE model. Thus, the full set of identifiable parameters and combinations for this model is
\[ \left\{ \mu_S, \mu_I, \frac{\beta}{k_I}, \beta c \epsilon,  \frac{\delta k_I}{k_E}, \frac{\delta(\epsilon-1)}{\epsilon},\mu_E+\delta \right\} \]
which is equivalent to the set of identifiable combinations for the ODE model.

\subsubsection{Identifiability of initial conditions}\label{sec:ic1}
With generic initial conditions, we do not gain any additional identifiability results. However, if we assume that certain information is known, such as the total population size at the initial time, then further model parameters can become identifiable. Here, we assume that the initial population size as well as its derivatives with respect to time and age are known for all ages. Under these conditions, we may solve for the initial susceptible population $S(0,a)$ as $S(0,a)=N(0,a)-E(0,a)-I(0,a)$, where $N$ represents the total population size. Similarly, since we assume the derivatives of $N$ are known at $t=0$, we may solve for the derivatives of $S$ at $t=0$ in terms of the derivatives of $E$ and $I$. It is thus straightforward to solve for $S$ and its derivatives at $t=0$ in terms of the outputs $y_E$ and $y_I$.

To determine parameter identifiability from the initial condition with this additional knowledge, we evaluate the model equations at $t=0$ and perform identifiability analysis analogous to the previous analysis for the boundary conditions. We obtain the following set of polynomial coefficients:
\begin{equation*}
\begin{aligned}
    \left\{1,\frac{k_I}{k_E},\frac{\beta }{k_E},\frac{\beta }{k_I},-\frac{\delta  k_I}{k_E},-\frac{\beta  k_E (\epsilon -1)}{k_I^2},\frac{\beta k_E (\epsilon -1)}{k_I},\frac{\beta  \epsilon }{k_E}, \right. \\
    \left. \frac{\beta  \epsilon }{k_I},\frac{\beta -\beta  \epsilon }{k_I},\delta +\mu_E,\mu_I-\beta  \epsilon ,\frac{k_I \mu_S}{k_E},\mu_S-\beta ,-k_I (\mu_S+2)\right\}
\end{aligned}
\end{equation*}
From these coefficients, we find that $\beta, \delta, \epsilon, \mu_S, \mu_E, \mu_I, k_E,$ and $k_I$ are all identifiable. Combining this with the previously established identifiability of $c\beta\epsilon$ from the boundary equations, we find that \emph{all parameters are identifiable} if the initial population size and its derivatives are known.

\section{Case study 2: Frequency-dependent transmission}

While density-dependent transmission is convenient for simplicity, it is not generally a realistic model for TB. Rather, most models for TB epidemiology use frequency-dependent transmission models wherein the transmission term is proportional to $S\frac{I}{N}$. In this section we present identifiability results for the SEI model with frequency-dependent transmission, assuming that our output functions are $y_E = k_E E$ and $y_I = k_I I$. In addition to the ODE and constant-parameter PDE models, we will also consider variations including immigration and age-dependent death rates. The full details of the analysis are cumbersome and thus are not provided here, but can be found in the Supplementary Material.

\subsection{ODE SEI model with frequency-dependent transmission} \label{sec:ODE-freq}
For completeness, we include here identifiability results for the ODE version of the model. This model can be written as the following system of equations:
\begin{equation}\label{eq:frequency_ODE}
\begin{aligned}
\frac{dS}{dt} &= c-\beta S\frac{I}{N} - \mu_S S \\
\frac{dE}{dt} &= (1-\epsilon)\beta S\frac{I}{N} - \delta E - \mu_E E \\
\frac{dI}{dt} &= \epsilon \beta S\frac{I}{N} + \delta E - \mu_I I
\end{aligned}
\end{equation}
where $N(t) = S(t)+E(t)+I(t)$. We find that the set of identifiable parameters and parameter combinations for this model is 
\[ \left\{ \beta, \delta, \epsilon, \mu_S, \mu_E, \mu_I, ck_I, \frac{k_E}{k_I} \right\}. \]

\subsection{Age-structured PDE SEI model with frequency-dependent transmission and constant parameters}
\label{sec:PDE_constant-freq}

\begin{equation} \label{eq:PDE_model-freq}
\begin{aligned}
\frac{\partial S}{\partial t}+\frac{\partial S}{\partial a} &= - \beta S\frac{I}{N} - \mu_S S \\
\frac{\partial E}{\partial t}+\frac{\partial E}{\partial a} &= (1-\epsilon)\beta S\frac{I}{N} - \delta E - \mu_E E \\
\frac{\partial I}{\partial t}+\frac{\partial I}{\partial a} &= \epsilon \beta S\frac{I}{N} + \delta E - \mu_I I
\end{aligned}
\end{equation}
with boundary conditions $S(t,0)=c$, $E(t,0)=I(t,0)=0$. From the model equations, we find that the parameters and combinations $\{ \beta, \delta, \epsilon, \mu_S, \mu_E, \mu_I, \frac{k_E}{k_I}\}$ are identifiable. We further find from the boundary equations that $c k_I$ is identifiable. Thus, the full set of identifiable parameters and combinations for this model is 
\[ \{ \beta, \delta, \epsilon, \mu_S, \mu_E, \mu_I, c k_I, \frac{k_E}{k_I}\} \]
which is consistent with the ODE model. If the initial age distribution of the total population and its derivatives were known (as in Section \ref{sec:ic1} for the previous case study), then all parameters in the model would be identifiable.

\subsection{SEI model with frequency-dependent transmission and immigration} \label{sec:immigration}
The models considered thus far have included influx of new individuals only via birth. However, it may be necessary to also consider immigration when studying epidemics in fluid populations. If immigration is included in the model, with the assumptions that the immigration rate is constant over time and all immigrants are susceptible, the ODE system becomes 
\begin{equation}\label{eq:migration_ODE}
\begin{aligned}
\frac{dS}{dt} &= m+c-\beta S\frac{I}{N} - \mu_S S \\
\frac{dE}{dt} &= (1-\epsilon)\beta S\frac{I}{N} - \delta E - \mu_E E \\
\frac{dI}{dt} &= \epsilon \beta S\frac{I}{N} + \delta E - \mu_I I
\end{aligned}
\end{equation}
where $m$ is the immigration rate. If we assume that the age distribution $\theta(a)$ of immigrants is known, then the corresponding age-structured PDE system becomes 
\begin{equation} \label{eq:migration_PDE}
\begin{aligned}
\frac{\partial S}{\partial t}+\frac{\partial S}{\partial a} &= m\theta(a) - \beta S\frac{I}{N} - \mu_S S \\
\frac{\partial E}{\partial t}+\frac{\partial E}{\partial a} &= (1-\epsilon)\beta S\frac{I}{N} - \delta E - \mu_E E \\
\frac{\partial I}{\partial t}+\frac{\partial I}{\partial a} &= \epsilon \beta S\frac{I}{N} + \delta E - \mu_I I
\end{aligned}
\end{equation}
where $\theta(a)$ is some known function such that $\theta(0)=0$ and $\int_0^\infty \theta(a) da = 1$, with boundary conditions $S(t,0)=c$ and $E(t,0)=I(t,0)=0$. 

The identifiability results for the ODE system \eqref{eq:migration_ODE} remain the same as before but with $c$ replaced by $c+m$, i.e., the identifiable parameters and combinations are
\[\left\{ \beta, \delta, \epsilon, \mu_S, \mu_E, \mu_I, (c+m)k_I, \frac{k_E}{k_I} \right\}. \]
For the PDE system \eqref{eq:migration_PDE}, we get the following identifiable parameters and combinations from the model and boundary equations:
\[\left\{ \beta, \delta, \epsilon, \mu_S, \mu_E, \mu_I, c k_I, m k_I, \frac{k_E}{k_I} \right\}. \]
Thus, in this case the identifiability results for the PDE model differ from those for the ODE model in that the combinations $c k_I$ and $m k_I$ are both identifiable, whereas in the ODE model only the sum $(c+m)k_I$ is identifiable. In other words, if all immigrants are susceptible then in an age-structured model it is possible to differentiate between birth and immigration, whereas in an ODE model it is not.

\subsection{Age-structured PDE SEI  model with frequency dependent transmission and  age-dependent parameters}

A major benefit of age-structured models is the ability to incorporate age-dependent parameters such as death rates, contact/transmission rates, and reactivation rates to make the model more realistic. Here, we consider the case of age-dependent death rates. There are several ways to introduce age dependence to the death rate: one could assume a specific functional form (e.g., piecewise-constant, linear, exponential), that the function belongs to a particular family (e.g., polynomials, splines), or that the function is completely arbitrary with some conditions on smoothness. In this section, we present identifiability results for four cases: (1) piecewise-constant death rate functions for all three compartments, (2) a single exponential death rate function for all compartments, (3) a single polynomial death rate function for all compartments, and (4) a single arbitrary function for all compartments. For simplicity, we will consider the SEI model without immigration.

\subsubsection{PDE SEI model with piecewise-constant death rates case}\label{sec:piecewise}
We first consider the case that the death rates are piecewise-constant functions of age with discontinuities at ages $\{0=a_0,a_1, a_2, ... a_n=A\}$, i.e. $\mu_j(a) = \mu_j^i$ for $a \in (a_i,a_{i+1})$ and $j=S,E,I$. This represents the case where death rates are known independently for different age groups. Then the system \eqref{eq:PDE_model-freq} becomes a sequence of PDE systems obeying the equations
\begin{equation*}
\begin{aligned}
\frac{\partial S}{\partial t}+\frac{\partial S}{\partial a} &= -\beta S\frac{I}{N} - \mu_S^i S \\
\frac{\partial E}{\partial t}+\frac{\partial E}{\partial a} &= (1-\epsilon)\beta S\frac{I}{N} - \delta E - \mu_E^i E \\
\frac{\partial I}{\partial t}+\frac{\partial I}{\partial a} &= \epsilon \beta S\frac{I}{N} + \delta E - \mu_I^i I
\end{aligned}
\end{equation*}
for $t>0$ and $a \in (a_i,a_{i+1})$, $0 \leq i < n$, with boundary conditions $S(t,0)=c$, $E(t,0)=I(t,0)=0$ and 
\begin{equation*}
    \begin{aligned}
    S(t,a_i)&=\lim_{a \rightarrow a_i^-} S(t,a) \\
    E(t,a_i)&=\lim_{a \rightarrow a_i^-} E(t,a) \\
    I(t,a_i)&=\lim_{a \rightarrow a_i^-} I(t,a)
    \end{aligned}
\end{equation*} 
for $0<i<n$. We will assume that the locations of the discontinuities $a_1,...,a_n$ are known, and that the sub-intervals are wide enough that the usual structural identifiability results hold.

Since the model equations remain essentially unchanged, structural identifiability of the model equations has already been established in Section \ref{sec:PDE_constant-freq}. Thus, from the model equations we obtain that 
\[ \{ \beta, \delta, \epsilon, \mu_S^i, \mu_E^i, \mu_I^i, \frac{k_E}{k_I}\} \] 
are identifiable for all $i$. From the boundary equations at $a=0$ we again get that $c k_I$ is identifiable, as in the constant parameter case. We gain no additional information about the parameter values from the boundary conditions at $a=a_i$ for $i>0$.

\subsubsection{PDE SEI model with exponential death rate case}\label{sec:exponential}
One may also wish to use a continuous function to represent age-dependent death rates. One possibility is to use an exponential function $\mu(a) = \mu_0 \exp(\kappa a)$ so that death rate increases exponentially with age. For simplicity, we assume that death rates are the same for all three compartments. Then the model system can be written as
\begin{equation*}
\begin{aligned}
\frac{\partial S}{\partial t}+\frac{\partial S}{\partial a} &= -\beta S\frac{I}{N} - \mu S \\
\frac{\partial E}{\partial t}+\frac{\partial E}{\partial a} &= (1-\epsilon)\beta S\frac{I}{N} - \delta E - \mu E \\
\frac{\partial I}{\partial t}+\frac{\partial I}{\partial a} &= \epsilon \beta S\frac{I}{N} + \delta E - \mu I \\
\frac{d\mu}{da} &= \kappa \mu
\end{aligned}
\end{equation*}
for $t>0$ and $a \in [0,\infty)$, with boundary conditions $S(t,0) = c$, $E(t,0)=I(t,0)=0$, and $\mu(0)=\mu_0$. We write the system in this way rather than directly inserting $\mu(a)=\mu_0 \exp(\kappa a)$ so that we maintain a differential polynomial form.

To obtain an input-output equation from the model equations, we use the $I$ equation to solve for $S$ and the $E$ equation to solve for $\mu(a)$ (note that these choices are somewhat arbitrary). Then by noting that $I=y_I/k_I$ and $E=y_E/k_E$ and substituting into the $S$ and $\mu$ equations, we obtain rational equations in terms of the outputs and their derivatives. Thus, by multiplying by a common denominator, we obtain differential polynomials. We find that $\left\{\beta, \delta, \epsilon, \kappa, \frac{k_E}{k_I} \right\}$ are the identifiable parameters and parameter combinations. 

By performing similar analysis on the boundary equations, we find that $\mu_0$ and $c k_I$ are identifiable. Thus, the full set of identifiable parameters and combinations for this version of the model is
\[ \left\{\beta, \delta, \epsilon, \kappa, \mu_0, c k_I, \frac{k_E}{k_I} \right\}. \]
This shows that both parameters from the exponential death rate are identifiable. The identifiability of the remaining parameters is consistent with the results for the ODE model in Section \ref{sec:ODE-freq}.

\subsubsection{PDE SEI model with polynomial death rate} \label{sec:polynomial}
Another option for a continuous death rate is to use a polynomial, $\mu(a) = \sum_{i=0}^n \kappa_i a^i$. For example, a polynomial may be fit to the available data for death rate by age group. Then the model equations become
\begin{equation*}
\begin{aligned}
\frac{\partial S}{\partial t}+\frac{\partial S}{\partial a} &= -\beta S\frac{I}{N} -\left(\sum_{i=0}^N \kappa_i a^i\right) S \\
\frac{\partial E}{\partial t}+\frac{\partial E}{\partial a} &= (1-\epsilon)\beta S\frac{I}{N} - \delta E - \left(\sum_{i=0}^N \kappa_i a^i\right) E \\
\frac{\partial I}{\partial t}+\frac{\partial I}{\partial a} &= \epsilon \beta S\frac{I}{N} + \delta E - \left(\sum_{i=0}^N \kappa_i a^i\right) I 
\end{aligned}
\end{equation*}
for $t>0$ and $a \in [0,\infty)$, with boundary conditions $S(t,0)=c$ and $E(t,0)=I(t,0)=0$. The age $a$ is considered to be observable. By using the $I$ equation to solve for $S$ and substituting into the remaining equations, we obtain a set of input-output equations that are differential polynomials. We find that the corresponding monic polynomial contains the monomial terms $\kappa_i a^i y_I$ for all $i=0,...,n$. Thus, all coefficient parameters $\kappa_i$ in $\mu(a)$ are identifiable and $\mu(a)$ can be uniquely recovered. We also find that $\left\{ \beta, \delta, \epsilon,\frac{k_E}{k_I} \right\}$ are identifiable from the model equations and $c k_I$ is identifiable from the boundary equations.

\subsubsection{PDE SEI model with arbitrary death rate function}\label{sec:arbitrary}
The final and most general option we consider is to allow $\mu(a)$ to be an arbitrary function of $a$, i.e. the model equations become
\begin{equation*}
\begin{aligned}
\frac{\partial S}{\partial t}+\frac{\partial S}{\partial a} &= -\beta S\frac{I}{N} - \mu(a) S \\
\frac{\partial E}{\partial t}+\frac{\partial E}{\partial a} &= (1-\epsilon)\beta S\frac{I}{N} - \delta E - \mu(a) E \\
\frac{\partial I}{\partial t}+\frac{\partial I}{\partial a} &= \epsilon \beta S\frac{I}{N} + \delta E - \mu(a) I
\end{aligned}
\end{equation*}
for $t>0$ and $a \in [0,\infty)$. In this case, we use the $I$ equation to solve for $S$ and then use the $E$ equation to solve for $\mu(a)$. We obtain the following expression for $\mu(a)$ in terms of the outputs and parameters:
\begin{equation} \label{eq:mu}
\begin{aligned}
\mu(a) &= -\frac{1}{k_E (\epsilon -1) y_I+k_I \epsilon  y_E}\left(\delta k_I y_E +k_I \epsilon  y_E^{(1,0)}  \right. \\
&\quad\left. +k_I \epsilon  y_E^{(0,1)}+k_E(\epsilon-1) y_I^{(1,0)}+k_E (\epsilon-1) y_I^{(0,1)}\right) \\
&= -\frac{1}{(\epsilon -1) y_I+\frac{k_I}{k_E} \epsilon  y_E}\left(\delta \frac{k_I}{k_E} y_E +\frac{k_I}{k_E} \epsilon  y_E^{(1,0)}  \right. \\
&\quad\left. +\frac{k_I}{k_E} \epsilon  y_E^{(0,1)}+(\epsilon-1) y_I^{(1,0)}+ (\epsilon-1) y_I^{(0,1)}\right) \\
\end{aligned}
\end{equation}
By substituting \eqref{eq:mu} into the remaining equation for $S$ and performing the identifiability analysis, we again find that $\left\{ \beta, \delta, \epsilon, \frac{k_I}{k_E} \right\}$ are identifiable. We note that the second derivatives of $y_E$ and $y_I$ appear in the input-output equations, and hence $\mu(a)$ must be at least twice differentiable. We then find that the right hand side of \eqref{eq:mu} consists entirely of observable quantities and identifiable parameter combinations; hence $\mu(a)$ is identifiable. Note that while we use the term ``identifiable" here, what we actually mean is that $\mu(a)$ can be determined from observable quantities and identifiable parameter combinations. Further, we obtain from the boundary equations that $c k_I$ is identifiable.

Identifiability results for all cases of the ODE and age-structured PDE model systems are summarized in Table \ref{tab:all}. We note that in the cases where the same death rate is used for all compartments, the death rate function is always identifiable. In cases where the compartments have different death rates, the rates for the susceptible and infected compartments are identifiable. These results hold under the assumption that a fraction of the total prevalence and a fraction of all exposures can be observed at all times and ages. It has been shown previously that other output functions may lead to non-identifiability of death rates even in simple population models \citep{Perasso16}.

\begin{table*}[b]
    \centering
    \def\arraystretch{1.5}
    \begin{tabular}{|c|c|} \hline
    \textbf{Case} & \textbf{Identifiable parameters and combinations} \\ \hline
    \makecell{ \\ Density-dependent ODE \\
    (Section \ref{sec:ODE-dens})\\} & $ \displaystyle \left\{ \mu_S, \mu_I, \frac{\beta}{k_I}, \beta c \epsilon,  \frac{\delta k_I}{k_E}, \frac{\delta(\epsilon-1)}{\epsilon},\mu_E+\delta \right\}$ \\
    \makecell{ \\  Density-dependent PDE with constant parameters \\ 
    (Section \ref{sec:PDE_constant-dens}) \\} & \makecell{$ \displaystyle \left\{ \mu_S, \mu_I, \frac{\beta}{k_I}, \beta c \epsilon,  \frac{\delta k_I}{k_E}, \frac{\delta(\epsilon-1)}{\epsilon},\mu_E+\delta \right\}$} \\
    \makecell{ \\ Frequency-dependent ODE \\
    (Section \ref{sec:ODE-freq})\\} & $ \displaystyle \left\{ \beta, \delta, \epsilon, \mu_S, \mu_E, \mu_I, ck_I, \frac{k_E}{k_I} \right\}$ \\
    \makecell{ \\  Frequency-dependent PDE with constant parameters \\ 
    (Section \ref{sec:PDE_constant-freq}) \\} & \makecell{$ \displaystyle \left\{ \beta, \delta, \epsilon, \mu_S, \mu_E, \mu_I, ck_I, \frac{k_E}{k_I} \right\}$} \\
    \makecell{ \\ Frequency-dependent ODE with immigration\\ (Section \ref{sec:immigration})\\ } & $ \displaystyle \left\{ \beta, \delta, \epsilon, \mu_S, \mu_E, \mu_I, (c+m)k_I, \frac{k_E}{k_I} \right\}$ \\
    \makecell{ \\ Frequency-dependent PDE with immigration\\ (Section \ref{sec:immigration})\\ } & $ \displaystyle \left\{ \beta, \delta, \epsilon, \mu_S, \mu_E, \mu_I, c k_I, m k_I, \frac{k_E}{k_I} \right\}$ \\
    \makecell{ \\ Frequency-dependent PDE with piecewise-constant death rates \\ (Section \ref{sec:piecewise})\\ } & $\displaystyle \left\{ \beta, \delta, \epsilon, ck_I, \frac{k_E}{k_I} \right\} \cup \left( \bigcup_{i=0}^{n-1} \{\mu_S^i, \mu_I^i, \mu_E^i \} \right)$\\
    \makecell{ \\ Frequency-dependent PDE with exponential death rate \\ (Section \ref{sec:exponential})\\ } & $ \displaystyle\left\{\beta, \delta, \epsilon, \kappa, \mu_0, c k_I, \frac{k_E}{k_I} \right\}$ \\
    \makecell{ \\ Frequency-dependent PDE with polynomial death rate \\ (Section \ref{sec:polynomial})\\ } &
    $ \displaystyle\left\{ \beta, \delta, \epsilon, ck_I, \frac{k_E}{k_I} \right\} \cup \left( \bigcup_{i=0}^{n} \{\kappa_i \} \right)$ \\
    \makecell{ \\ Frequency-dependent PDE with arbitrary death rate function \\ (Section \ref{sec:arbitrary})\\ \\ } &
    $ \displaystyle\left\{ \beta, \delta, \epsilon, ck_I, \frac{k_E}{k_I}, \mu(a) \right\}$ \\ \hline\end{tabular}
    \caption{Summary of identifiable parameters and combinations for the ODE and age-structured PDE systems where the output functions are $y_E = k_E E$ and $y_I = k_I I$.}
    \label{tab:all}
\end{table*}

\subsection{Practical identifiability}
The above structural analysis determines identifiable parameters and combinations under the assumption that data are noiseless and are available for arbitrarily many time and age points. In reality, epidemiological data are often noisy and are collected only at discrete time points and for discrete age groups. Thus, it is also necessary to perform practical identifiability analysis to determine which parameters and parameter combinations are still identifiable given more realistic data. The practical identifiability of model parameters given realistic data may differ from their structural identifiability.

As a proof of concept, we explore the practical identifiability of the constant-parameter model from Section \ref{sec:PDE_constant-freq} using profile likelihoods. We assume that  data are collected monthly for 5-year age groups from age 0 to 100 over a span of 20 years. We synthetically generate data by choosing a baseline parameter set, simulating the model, and adding noise that is normally distributed with a standard deviation of 5\%. Using this synthetic data, we perform parameter estimation using the Matlab function {\it fminsearch}, and evaluate practical identifiability of the model parameters as detailed in Section \ref{sec:practical}.

\begin{figure}
    \centering
    \includegraphics[width=0.7\textwidth]{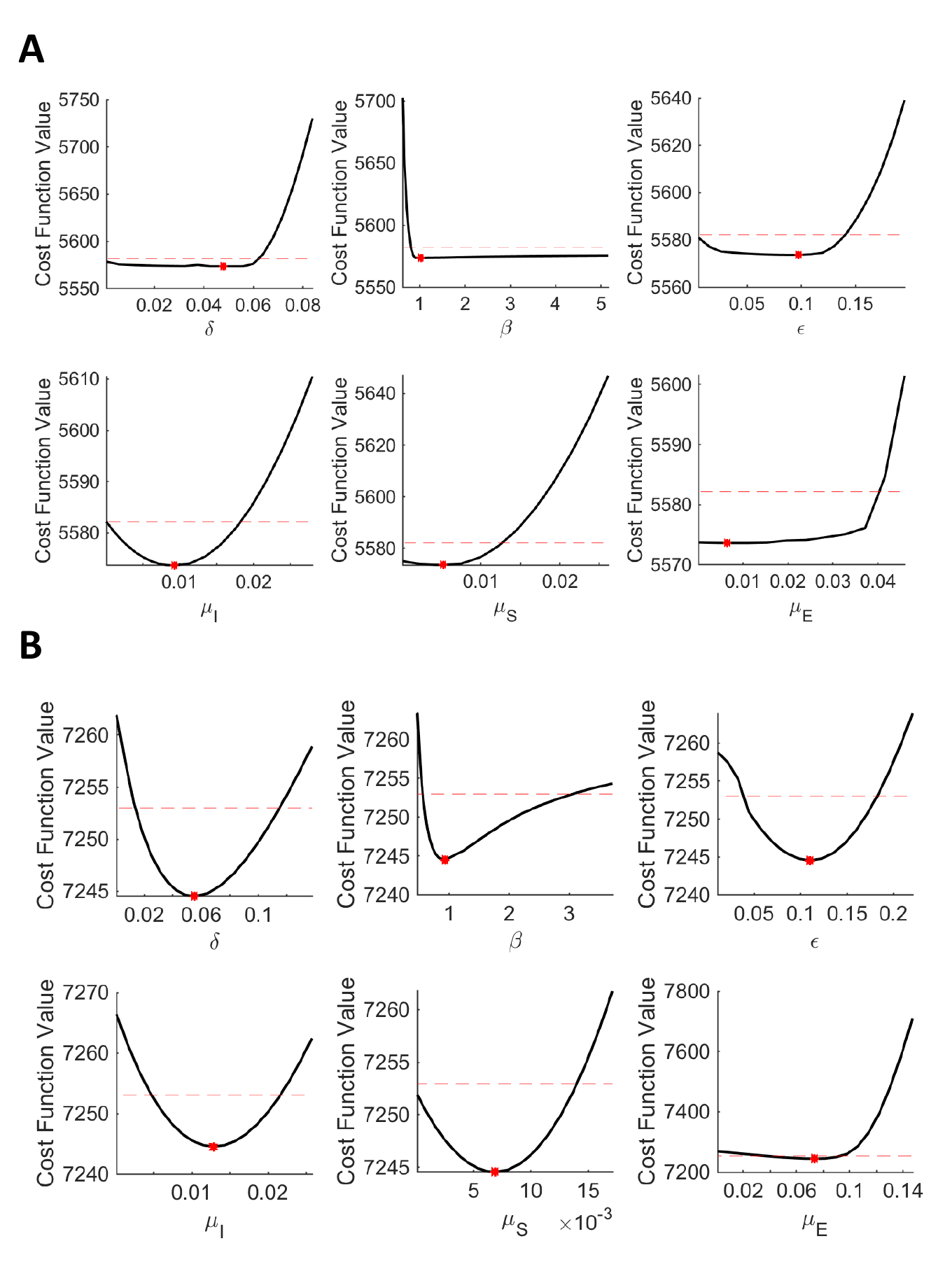}
    \caption{Practical identifiability of the constant-parameter model from Section \ref{sec:PDE_constant-freq}. Cost functions (black lines) based on likelihoods are shown for two sets of parameter values, $P_1 = \{c=100, \beta=1, \epsilon=0.1, \delta=0.05, \mu_S=0.005, \mu_E=0.005, \mu_I=0.01, k_E=0.5, k_I=0.5\}$ (panel A) and $P_2 = \{c=100, \beta=1, \epsilon=0.1, \delta=0.05, \mu_S=0.005, \mu_E=0.075, \mu_I=0.01, k_E=0.75, k_I=1\}$ (panel B). Red dashed lines denote 95\% confidence thresholds, and red dots denote local minima (corresponding to maximum likelihoods). See Section \ref{sec:practical} for more details of the analysis.}
    \label{fig:practical}
\end{figure}

We compute the Fisher Information Matrix (FIM) using the outputs $y_E$ and $y_I$ at the above-specified ages and time points. We find that for a wide variety of parameter sets within reasonable ranges, the rank of the FIM is eight, implying that there should be eight identifiable parameters and parameter combinations. This is consistent with our structural identifiability analysis performed in Section \ref{sec:PDE_constant-freq}, which revealed a set of eight identifiable parameters and combinations. From the structural analysis, we see that all parameters except $c, k_I$, and $k_E$ are structurally identifiable. However, from profile likelihoods we observe that the practical identifiability of these parameters does vary across the parameter space. Here, we demonstrate two baseline parameter sets that lead to qualitatively different identifiability profiles (Figure \ref{fig:practical}). For both parameter sets, $c, k_E,$ and $k_I$ are not practically identifiable.

Using the parameter set $P_1 = \{c=100, \beta=1, \epsilon=0.1, \delta=0.05, \mu_S=0.005, \mu_E=0.005, \mu_I=0.01, k_E=0.5, k_I=0.5\}$, we find that certain structurally identifiable parameters are only practically identifiable on one side, that is, we can establish a reasonable lower or upper bound but not both. While the profile likelihoods do have a unique minimum, the profiles are flat and confidence intervals for these parameters are thus wide. In particular, the transmission rate $\beta$ cannot be reasonably bounded from above, and $\epsilon, \delta, \mu_S,$ and $\mu_E$ cannot be distinguished from zero. 

Using a slightly different parameter set $P_2 = \{c=100, \beta=1, \epsilon=0.1, \delta=0.05, \mu_S=0.005, \mu_E=0.075, \mu_I=0.01, k_E=0.75, k_I=1\}$, in which reporting rates and death among the exposed population are increased, we find that all structurally identifiable parameters are also practically identifiable. While this parameter set is not necessarily biologically realistic (since $\mu_E > \mu_I$, for example), this demonstrates that there are regions in the parameter space where these parameters can be well estimated.

\section{Summary and Discussion}
Creating mathematical and computational models can greatly impact the biological community through predictions that can be used and tested. Parameterization of models is a key factor in connecting them to the biology, and thus identifying parameters accurately is an important step of model development. Parameters are typically estimated from data; however, data may be sparse and noisy, and even perfect data does not guarantee unique estimability of parameters. Understanding if model parameters are identifiable lends credibility to the model system. If parameters are not identifiable, then the applicability  of those models becomes more theoretical. While this may add value to understanding, it limits the potential for predictions to be translated in a meaningful way toward impacting the biological system. To this end, we focus on parameter identifiability in both ODE and PDE models. While there is a solid literature on structural identifiability analysis for ODE models, there is not a unified framework for studying identifiability in PDEs. Here, we have presented a methodology for performing structural identifiability analysis on age-structured epidemic models using a differential algebra framework. We have applied this methodology to explore the structural identifiability of a simple age-structured Susceptible-Exposed-Infected (SEI) epidemic model under various assumptions. 

We found that when the age-structured model uses constant parameters and does not include immigration, then the identifiability results are identical to those for the corresponding ODE model. In the presence of immigration, however, differences in identifiability arise: the rates for birth and immigration appear only as a sum in the identifiable combinations of the ODE model, while they appear separately in those for the age-structured PDE model formulation. Further, we found that age-dependent death parameters in the PDE model are able to be uniquely recovered in at least the following three cases: piecewise-constant death rates, a single exponential death rate for all compartments, and a single polynomial death rate for all compartments.

These findings imply that identifiability results for corresponding ODE and PDE models are likely equivalent when parameters are constant, but that additional information can be gained from PDE models when parameters vary with the non-temporal variable(s). Rigorous derivation of necessary and sufficient conditions for different models to have identical identifiability results remains to be done, however based on our results herein, we conjecture that identifiability for the ODE and PDE forms of an age-structured model should be equivalent when there are not age-dependent parameters present in the model. Additionally, future work to investigate how discretization of the PDE (e.g. converting age dynamics into linear `chains' of compartments for age groups \citep{hurtado2019generalizations,hurtado2020time}) impacts identifiability of the model would be a useful direction given the common use of compartmental representations of aging in biological models.

The framework presented here could be extended to other PDE models, such as reaction-diffusion and advection-diffusion equations, in a variety of application areas. In particular, the assumption used here that age-interactions are only local (i.e. individuals interact with only people their own age) may be reasonable for reaction-diffusion models. However, since this methodology relies on the ability to express the input-output equations as differential polynomials, there is a limit to the amount of complexity that it can handle. Thus, highly nonlinear models and/or output functions may require the use of other tools such as numerical methods. One such method is the FIM, which we have demonstrated in the context of an age-structured model with preferential mixing informed by real data.

A significant limitation of structural identifiability analysis is that it assumes the availability of perfect data. In practice, data may contain noise and bias, and may be sampled only at discrete time points. Thus, practical identifiability analysis must also be performed to gain a complete understanding of the parameter identifiability of a model given the available data. As demonstrated in our examples, practical identifiability of a model may vary not only with the quality of the data but also with the corresponding values of model parameters. Practical identifiability thus needs to be handled on a case-by-case basis for models and their corresponding parameter sets. Taken together, practical and structural identifiability provide rigor to the modeling process, adding to the applicability of model predictions.

\section*{Acknowledgements}
This research was supported by NIH grants R01AI123093 and U01HL131072 awarded to DK and NSF DMS Grant 1853032 and NIH MIDAS Grant U01 GM110712 awarded to ME. 

\bibliographystyle{spbasic}
\bibliography{references}

\end{document}